\documentstyle[prl,aps,preprint,psfig]{revtex}
\tightenlines
\newlength{\figwidth}
\title{The Riemannium}
\author{P. Leboeuf, A. G. Monastra, and O. Bohigas}
\address{Laboratoire de Physique Th\'eorique et Mod\`eles Statistiques,
B\^at. 100, \\ Universit\'e de Paris-Sud, 91405 Orsay Cedex, France}
\begin{document}
\maketitle {\begin{abstract} The properties of a fictitious, fermionic,
many-body system based on the complex zeros of the Riemann zeta function are
studied. The imaginary part of the zeros are interpreted as
mean-field single-particle energies, and one fills them up to a Fermi energy
$E_F$. The distribution of the total energy is shown to be non-Gaussian,
asymmetric, and independent of $E_F$ in the limit $E_F\rightarrow\infty$. The
moments of the limit distribution are computed analytically. The
autocorrelation function, the finite energy corrections, and a comparison with
random matrix theory are also discussed.
\end{abstract}}
{}
\narrowtext

According to the Riemann hypothesis, the complex zeros of the function $\zeta
(s) = \sum n^{-s}$ (defined for ${\rm Re} \, s>1$, and by analytic
continuation to the rest of the complex plane) are located on the critical
line $s_\mu = 1/2 + i \, E_\mu$, with $E_\mu$ real. There is a natural
connexion between this hypothesis and quantum mechanics that originates on a
spectral interpretation of the complex zeros $s_\mu$. Indeed, a general
strategy to prove the hypothesis was suggested by Hilbert and P\'olya, who
proposed to find an hermitian operator whose eigenvalues are precisely the
$E_\mu$'s. This operator may be viewed as the quantization of an hypothetical
classical dynamical system, which determines the ``Riemann dynamics''.
Although the classical Hamiltonian corresponding to the Riemann dynamics is
not known, there are evidences for its existence. When the $E_\mu$ are
interpreted as quantum eigenvalues, both their statistical properties
\cite{mont,odlyzko,bgs} as well as a semiclassical interpretation of their
density \cite{berry} indicate that the Riemann dynamics is fully chaotic and
has no time-reversal symmetry. The semiclassical interpretation provides in
fact much more detailed information (cf Eq.(\ref{2})), in particular
concerning the classical periodic orbits. This leads to a somewhat paradoxical
situation: although the ``Riemann Hamiltonian'' is not known, the detailed
properties of its dynamics are. From the perspective of a dynamicist, the
Riemann zeta thus offers the rare opportunity of a chaotic motion for which
the relevant dynamical information is simple and explicitly known. Not to
mention the impressive amount of existing numerical data on the complex zeros
\cite{odlyzko}, as well as the number-theoretic background of $\zeta (s)$,
also useful in the present context.

Many aspects of the Riemann dynamics have been investigated (see \cite{sar,bk}
for recent review articles). Here we explore new facets of the problem related
to a different use of the complex zeros as suggested by physical analogies. We
consider the location $E_\mu >0$ of the complex zeros of $\zeta (s)$ as the
single-particle levels of a fermionic many-body system. In the mean field
approximation, the ground-state total energy is obtained by filling the
single-particle levels from the lowest Riemann zero up to a ``Fermi energy''
$E_F$. We are interested in the properties of such a Fermi gas. Following
nuclear physics terminology, we refer to this ``element'' as the Riemannium.

At zero temperature, the properties of the Riemannium are described by the 
grand potential
\begin{equation}\label{1}
\Omega (E_F) = \int_0^{E_F} \!\! (E-E_F) \rho (E) dE = 
- \int^{E_F}_{0} \!\! {\cal N} (E) dE \ .
\end{equation}
Here $\rho (E) = \sum _\mu \delta (E - E_\mu)$ is the spectral density of the
complex Riemann zeros interpreted as quantum eigenvalues and 
${\cal N} (E) = \int \rho (E) dE$ its counting function. $\Omega (E)$ 
corresponds to the sum of the imaginary part of the Riemann zeros satisfying 
$0 < E_\mu \leq E_F$, using $E_F$ as reference energy \cite{bls1}.
It is therefore the total energy of the system. To calculate the grand
potential we make use of the decomposition of the spectral density in smooth
plus oscillatory parts, $\rho = \overline{\rho} + \widetilde{\rho}$. The
former, as a Weyl series, has an explicit asymptotic expansion for large $E$.
The oscillating term is an interference sum over the prime numbers
$p=2,3,\ldots$ \cite{tit}
\begin{equation}\label{2}
\widetilde{\rho} (E) = - \frac{1}{\pi} {\rm Re} \sum_p\sum_{r=1}^\infty 
\frac{\log p}{p^{r/2}} \exp (i \, r E \log p) \ .
\end{equation}
The comparison of this equation with the semiclassical Gutzwiller trace
formula for the spectral density of a dynamical system \cite{gutz},
$\widetilde{\rho}(E) = \sum_{po} A_{po} \cos (S_{po}(E)/\hbar)$ where $po$ are
the classical periodic orbits, shows that each prime number labels an unstable
periodic orbit of a fully chaotic system of action $S_p = E \, T_p$, period
$T_p = r \log p$, Lyapounov stability $\lambda_p = 1$, repetitions labeled by
$r$, and $\hbar=1$ \cite{berry}. An unusual fact of the Riemann dynamics that
plays an important role in what follows is the independence with respect to
the energy $E$ of the periods and Lyapounov exponents (notice also the minus
sign in front of Eq.(\ref{2})).

The spectral density can be integrated twice to obtain the smooth and
oscillatory contributions to the grand potential, $\overline{{\Omega}} (E_F) =
-E_F^2 \log (E_F/2 \pi)/4 \pi + 3 E_F^2/8 \pi - 7 E_F/8 - \log E_F/48 \pi + 
cte$, and
\begin{equation}\label{3}
\widetilde{{\Omega}} (E_F) = - \frac{1}{\pi} {\rm Re} \sum_p
\sum_{r=1}^{\infty} \frac{\exp(i \, E_F \, r \log p)}{r^2 ~ p^{r/2} \log p}  \ .
\end{equation}
We are interested in the statistical properties of $\widetilde{{\Omega}}$ as a
function of the Fermi energy $E_F$. From Eq.(\ref{3}) we have
$\langle\widetilde{{\Omega}}\rangle=0$. The average is done over an energy
window which is small compared to $E_F$ but contains several oscillations of
$\widetilde{{\Omega}}$ (the typical scale of oscillation will be given below).

As we will now see, the Riemannium has very peculiar properties. The most
important ones demonstrated here are: i) the distribution of
$\widetilde{\Omega}$, denoted $P(\widetilde{{\Omega}})$, is independent of
energy as $E_F \rightarrow\infty$; ii) $P(\widetilde{{\Omega}})$ is
non-Gaussian and asymmetric; iii) all the moments of
$P(\widetilde{{\Omega}})$ may be computed with very good accuracy from
Eq.(\ref{3}), their value being dominated by the contribution of the lowest
prime numbers; iv) the asymmetry of the distribution is due to an interference
effect between repetitions of periodic orbits; v) the autocorrelation
$\langle\widetilde{{\Omega}} (E_F) \widetilde{{\Omega}}(E_F +
\epsilon)\rangle$ is a non-decaying irregular oscillatory function; vi) the
finite energy corrections to $P(\widetilde{{\Omega}})$ are universal and well
described by a circular unitary ensemble of random matrices ($CUE$).

Some of the statistical properties of $\widetilde{{\Omega}}$ were explored in
the past. Selberg \cite{selberg} computed the even moments of the
distribution. Up to an additive constant and a global change of sign,
$\widetilde{{\Omega}} (E)$ coincides with the function denoted $S_1 (t)$ by
him. More recently, Odlyzko \cite{odlyzko} has numerically calculated the
first four moments, but no comparison has been made with the analytical
results of Selberg.

On Fig.~1 is displayed the distribution $P(\widetilde{{\Omega}})$ whose
properties are now discussed. We begin by computing the second moment. Equation
(\ref{3}) allows to express it as an integral over the period of the orbits
\begin{eqnarray} \label{4}
&&\langle\widetilde{{\Omega}}^2\rangle= (1/2\pi^2) \int_0^\infty dT \, K(T) / \, T^4 \ , \\
K(T) = && \langle \sum_{i,j} A_i A_j \cos[E_F (r_i \log p_i - r_j \log p_j)] 
\delta(T - {\bar T})\rangle. \nonumber
\end{eqnarray}
Use has been made here of the definition of the form factor, $K(T) \!=\!
4\pi\int_0^\infty d\epsilon \cos(\epsilon T) \langle \widetilde{\rho} (E_F)
\widetilde{\rho} (E_F+\epsilon) \rangle$, with the amplitudes defined as $A_i
= \log p_i/ p_i^{r_i/2}$, and ${\bar T} = (T_{p_i}+ T_{p_j})/2$. Rigorous
arguments valid for times $T_{min} \ll T \ll T_H$, where $T_{min} = \log 2$ is
the shortest period of the system and $T_H = 2 \pi {\bar \rho} =
\log(E_F/2\pi)$ is the Heisenberg time at $E_F$, and heuristic for longer
ones, show \cite{mont} that the form factor of the complex zeros of $\zeta
(s)$ tends, as $E_F$ goes to infinity, to the corresponding function $K_{GUE}$
of Gaussian random matrices with unitary symmetry (GUE). The latter behaves as
$K_{GUE} (T) = T$ for $T < T_H$, and $K_{GUE} (T) = T_H$ for $T > T_H$.
However, the replacement $K=K_{GUE}$ in Eq.(\ref{4}) and its extrapolation to
short times is of no use to understand the behavior of $\widetilde{{\Omega}}$,
because the integral diverges. In other words, the second moment is dominated
by the (non-universal) short periodic orbits whose contribution has to be
considered explicitly. We therefore keep as leading-order approximation the
diagonal part $(i=j)$ in the double sum (\ref{4}). This gives the {\sl
convergent} sum
\begin{equation} \label{5}
\langle\widetilde{\Omega}_0^2\rangle= \frac{1}{2\pi^2} \sum_p 
\sum_{r=1}^\infty \frac{1}{r^4 p^r \log^2 p} \approx 7.9\times 10^{-2} \ .
\end{equation}
This equation exhibits two of the main properties of
$P(\widetilde{{\Omega}})$. The first one is its asymptotic independence with
respect to energy. This fact is not generic for dynamical systems, since for
example for a Fermi gas in a chaotic cavity
$\langle\widetilde{{\Omega}}^2\rangle$ grows linearly with the Fermi energy
\cite{lm}. It is due to the independence of the periods and Lyapounov
exponents with respect to energy in the Riemann dynamics. The second property
is its non-universality (the sum (\ref{5}) depends on the particular
properties of the short orbits). The longer, statistically universal, periodic
orbits provide next-to-leading order corrections to Eq.(\ref{5}). For times $T
\gg T_{min}$ the form factor $K_{GUE}$ can be used in Eq.(\ref{4}), leading 
to the correction (independent of the prime numbers)
\begin{equation} \label{6}
\langle\widetilde{{\Omega}}^2\rangle =
\langle\widetilde{\Omega}_0^2\rangle - \frac{1}{12 \pi^2 \log^2
(E_F/2\pi)} \ .
\end{equation}
We have numerically checked the validity of Eq.(\ref{6}) down to values of
$E_F$ of the order of a few thousands.

The third basic feature of the distribution is its asymmetry, as revealed by
the third moment $\langle\widetilde{{\Omega}}^3\rangle$, whose computation is
now sketched. The third power of $\widetilde{{\Omega}}$ computed from
Eq.(\ref{3}) involves products of three cosines containing as argument the
action of three different prime numbers. This product may be expressed as a
sum of cosines involving the sum and differences of the actions. The term
involving the sum of actions has zero average. As before, the remaining sum
is dominated by the smallest primes. Due to the non-commensurability of the
periods of the different primes we restrict moreover to the approximation $p_i
= p_j = p_k = p$. The oscillating factors in the sum now have the typical form
$\cos [E (r_i + r_j - r_k) \log p]$. The only terms of this type having a
non-zero average are those satisfying $r_k = r_i + r_j$. Since there are three
different possibilities for choosing the backward repetition $r_k$, we finally
obtain the convergent sum
\begin{eqnarray} \label{7}
\langle\widetilde{{\Omega}}_0^3\rangle &&= -\frac{3}{4\pi^3} \sum_p 
\sum_{r_i,r_j=1}^\infty 
[ r_i^2 r_j^2 (r_i + r_j)^2  p^{r_i + r_j} \log^3 p ]^{-1} \nonumber \\
&& \approx - 5.78\times 10^{-3} \ .
\end{eqnarray}
The asymmetry of $P(\widetilde{{\Omega}})$ is therefore related to a simple
but non-trivial interference phenomenon occurring between two (different or
not) repetitions of a given orbit compensated by a backward one. The extreme
case where only $p=2$ and its repetitions are taken into account in
Eq.(\ref{7}) yields already $90\%$ of its value.

The higher moments of the distribution are computed likewise. In the limit
$E_F\rightarrow\infty$ to leading order they are given by expressions similar
to, though more complicated than, Eq.(\ref{7}). Table~I shows a comparison of
the analytical results with numerical data up to the sixth moment. The
agreement between both calculations is astonishing. Our numerical results also
coincide for the lowest moments with those obtained in \cite{odlyzko}.

An additional characterization of the Riemannium comes from the
autocorrelation of the total energy ${\cal C}_{\Omega} (\epsilon) = \langle
\widetilde{{\Omega}} (E_F) ~ \widetilde{{\Omega}} (E_F+\epsilon) \rangle/
\langle\widetilde{{\Omega}}_0^2\rangle$. We evaluate this quantity to leading
order by using the expansion (\ref{3}) and making a diagonal approximation.
This leads to
\begin{equation}\label{8} 
{\cal C}_{\Omega} (\epsilon)  =
\frac{1}{2 \pi^2 \langle\widetilde{{\Omega}}_0^2\rangle} \sum_p 
\sum_{r=1}^{\infty} \frac{\cos (\epsilon ~ r \log p)}{r^4 ~ p^r \log^2 p} \ .
\end{equation}
Due to the dominance of the shortest orbits in the sum (\ref{8}) ${\cal
C}_{\Omega}$ is an irregularly fluctuating non-decaying function, as
illustrated on Fig.~2. The fundamental period $2\pi/\log 2\approx 9.06$
associated to the shortest orbit can be clearly observed.

It is interesting to make a comparison of the previous results with random
matrix theory. As already emphasized, the $GUE$ is not really appropriate
because no short-time scale equivalent to $T_{min}$ is built-in in the theory
(and this produces a divergence of the moments). We therefore concentrate on
the $CUE$ which, contrary to the Gaussian one, has
an inherent short-time scale (although more appropriate measures, not relevant
for the present discussion, have been recently proposed \cite{kzs}). The
following analysis reveals also some striking similarities existing between
the Riemann zeros and eigenvalues of circular ensembles. Consider an $N\times
N$ unitary matrix $U$ describing the time-periodic dynamical evolution of a
quantum system. We fix for simplicity the periodicity to one. The Floquet
spectrum of $U$ is given by the eigenvalue equation $U \psi_\alpha = \exp(i
\theta_\alpha) \psi_\alpha$, $\alpha = 1, \ldots , N$. $\psi_\alpha$ are the
(stroboscopic) eigenstates and $\theta_\alpha$ the eigenphases. The spectral
density on the unit circle is $\rho_{CUE} (\theta) \!=\! \sum_\alpha
\delta(\theta - \theta_\alpha)$. The $2 \pi$--periodicity of this function
leads to a decomposition in smooth plus oscillatory parts, $\rho_{CUE} \!=\!
N/2\pi + \widetilde{\rho}_{CUE}$, with
\begin{equation}\label{9}
\widetilde{\rho}_{CUE} (\theta) = \frac{1}{\pi} {\rm Re} \sum_{k=1}^\infty 
{\rm Tr} U^k \exp (- i k \theta) \ .
\end{equation}
As was done for Eq.(\ref{2}), a direct comparison of Eq.(\ref{9}) with
Gutzwiller trace formula allows to make a ``na\"{\i}ve'' semiclassical interpretation of this equation. We look at it as a sum over the
periodic orbits of a classical map labelled by the index $k=1, 2, \ldots$,
having period $T_k = k$, action $S_k = k \ \theta$, (local) energy $\theta$,
stability amplitude $A_k = {\rm Tr} U^k$, and $\hbar=1$ (repetitions are
degenerated with the fundamental periods). The interpretation of $\theta$ as
the energy satisfies, as it should, the classical relation $T_k = \partial
S_k/\partial \theta$. We recover in this analogy one of the basic properties of
the Riemann dynamics: the independence of the periods with respect to energy.
Short orbits correspond to $T_k \approx T_{min} = 1$, while long ones
contributing to the density at the scale of the mean level spacing have
period $T_k \approx T_H = 2 \pi {\bar \rho} = N$. If now an ensemble of
unitary matrices is considered, Eq.(\ref{9}) acquires a statistical meaning
since the prefactors ${\rm Tr} U^k$ have now a distribution. It is well known
through, e.g., the prime number theorem and the Hardy-Littlewood conjecture, that asymptotically the statistical properties of large prime numbers are those required to mimic random matrix statistics \cite{mont,bk}. So for long
times Eqs.(\ref{2}) and (\ref{9}) are very similar from a statistical point of
view. This is not the case for times $T \approx T_{min}$. To have a
quantitative description we look at the $CUE$ distribution of the
grand potential, obtained by integrating $\widetilde{\rho}_{CUE}$ twice with
respect to $\theta$,
\begin{equation} \label{10}
\widetilde{\Omega}_{CUE} = (1/\pi) {\rm Re} \sum_k ({\rm Tr} U^k /k^2) 
exp(- i k \theta) \ .
\end{equation}
The second moment is 
\begin{equation}\label{11}
\langle \widetilde{\Omega}_{CUE}^2 \rangle \!\!=\!\!
(1/2\pi^2) \! \sum_k \langle |{\rm Tr} U^k|^2 \rangle /k^4 \ .
\end{equation} 
The average is done over the ensemble of matrices. This expression is the
$CUE$ analog of Eq.(\ref{4}) ($K_{CUE} = \langle |{\rm Tr} U^k|^2 \rangle$).
Using that $\langle |{\rm Tr} U^k|^2 \rangle = k$ if $k \leq N$ and $\langle
|{\rm Tr} U^k|^2 \rangle = N$ if $k > N$ \cite{hkssz}, we get
\begin{equation} \label{12}
\langle\widetilde{{\Omega}}_{CUE}^2\rangle = \frac{\zeta (3)}{2 \pi^2} - 
\frac{1}{12 \pi^2 N^2} + {\cal O} (1/N^4) \ ,
\end{equation}
to be compared to Eq.(\ref{6}). As for the Riemannium, the leading term $\zeta
(3)/2 \pi^2 \approx 6.1\times10^{-2}$ is controlled by short times (small
values of $k$) in the sum (\ref{11}). Because the short-time structure differ
in both cases, the constants do not coincide. But the independence of the
leading terms with respect to $N$ (or energy) is not a coincidence and comes
from the structural similarities between Eqs.(\ref{2}) and (\ref{9}).
Moreover, the next-to-leading order corrections are the same if one identifies
\cite{ks} the corresponding Heisenberg times $N=\log (E_F /2\pi)$. Both
corrections come from times of order $T_H$, where the statistical properties
agree. This situation is in contrast with the results obtained, for instance,
for the distribution of $\log \zeta (1/2 + i E)$, which has been shown to
agree to leading order with the (finite $N$) $CUE$ distribution, followed by
non-universal corrections \cite{ks}.

The qualitative agreement between $\langle\widetilde{{\Omega}}^2\rangle$ and
$\langle\widetilde{{\Omega}}_{CUE}^2\rangle$ is lost when considering odd
powers of $\widetilde{{\Omega}}_{CUE}$. The latter can be shown to vanish in the limit $N\rightarrow\infty$. More generally, and contrary to the Riemann case, in that limit the distribution of $\widetilde{\Omega}_{CUE}$ tends to a
Gaussian. This can be seen from Eq.(\ref{10}). On the one hand we have
demonstrated that $\widetilde{\Omega}_{CUE}$ is asymptotically dominated by
the lowest contributions in the sum. But ${\rm Tr} U^k$ is known to be, for
$k$ finite and $N\rightarrow\infty$, Gaussian independent distributed
\cite{hkssz}. Therefore $\widetilde{\Omega}$ has also a limiting normal
distribution.

The techniques employed here to analyze the distribution of the grand
potential are not specific to the Riemann dynamics and can be applied in a
wider context to the theory of Fermi gases. Our results indicate that in
general the grand potential of a Fermi gas, as well as other physical
quantities derived from it, are controlled by the shortest non-universal
classical orbits. This dominance leads, via the interference mechanisms
illustrated here for the grand potential of the Riemannium, to non-Gaussian
asymmetric distributions for the associated quantity. Some related results
illustrating this general phenomenon were obtained by Tsang \cite{tsang}, who
demonstrated non-Gaussian asymmetric distributions for the error term in the
mean square formula of $|\zeta(1/2 + i \, E)|$, and for the fluctuations of
the number of lattice points inside a curve, namely a circle (see also
\cite{bdl}) and an hyperbola (known as the circle problem and Dirichlet's
divisor problem, respectively). The last two problems have a dynamical
interpretation in terms of the fluctuations of the spectral counting function
of an associated integrable system. The mechanism leading to the asymmetry in
\cite{tsang} is very similar to the one found here. In contrast, the
fluctuations of the counting function in the Riemann case are known to be
asymptotically Gaussian \cite{selberg} and are not dominated by the short
orbits. This difference in the behavior of the counting function can be traced
back to the different short-time behavior of the form factor for integrable
and chaotic systems, that produces a dominance of the short orbits in the
former. We find however that integrals of the counting function (the grand
potential being the first one, cf Eq.(\ref{1})) are {\sl always} dominated by
the short orbits, irrespective of the (chaotic or integrable) nature of the
underlying classical dynamics. On the mathematical side, the present results
can be extended in a straightforward manner to general Dirichlet's
$L$-functions.

In summary we have considered, guided by physical analogies, new properties of
the Riemann zeros that reveal new aspects of their dynamical interpretation as
quantum eigenvalues of a classically chaotic system. Our results are relevant
in the theory of Fermi systems, as well as in the general context of quantum
chaotic motion. Concerning the Riemann hypothesis and the search of the
Hilbert--P\'olya Hamiltonian, the present study suggests that time--periodic
dynamical evolutions have to be considered as serious candidates.

We are specially grateful to D. Hejhal, J. Keating and A. Odlyzko for fruitful
discussions and suggestions. The Laboratoire de Physique Th\'eorique et
Mod\`eles Statistiques is a Unit\'e Mixte de Recherche de l'Universit\'e
Paris XI and CNRS.

\begin{figure}
\centerline{\psfig{figure=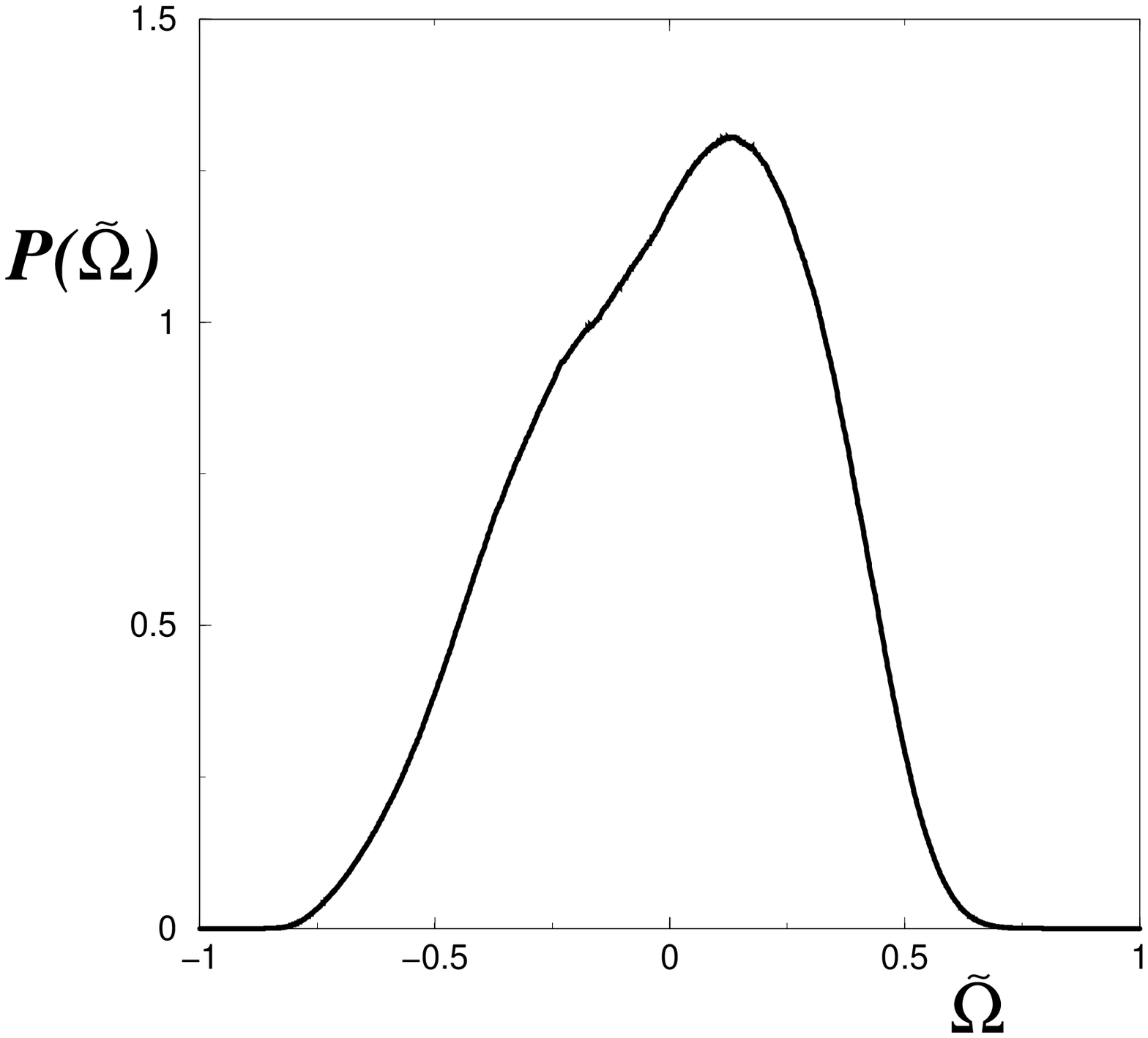,width=2.8in,height=2.2in}}
\caption{Distribution of $\widetilde{{\Omega}}$ computed numerically at
$E_F\approx 1.44\times 10^{20}$ (results based on data from A. Odlyzko).}
\end{figure}

\begin{figure}
\centerline{\psfig{figure=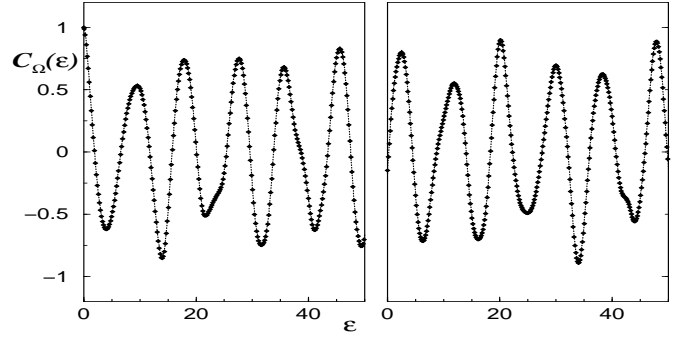,width=3.4in,height=1.8in}}
\caption{The autocorrelation Eq.(\ref{8}) (dotted line) compared to numerical
data (dots). $E_F$ as in Fig.~1. On the right part a constant $\epsilon_0
\approx 1.29\times10^{20}$ should be added to the abscissa.}
\end{figure}

\begin{table}[b]
\begin{tabular}{||r|r|r||}
Moment & Semiclassics~~~~~~~~ & Numerics~~~~~~~~ \\
\hline
$2$~~~~ & $7.9290\times 10^{-2}$~~~~~~  & $7.928\times 10^{-2}$~~~~~ \\
$3$~~~~ & $-5.7822\times 10^{-3}$~~~~~~ & $-5.785 \times 10^{-3}$~~~~~ \\
$4$~~~~ & $1.4814\times 10^{-2}$~~~~~~  & $1.481\times 10^{-2}$~~~~~ \\
$5$~~~~ & $-2.7787\times 10^{-3}$~~~~~~ & $-2.776\times 10^{-3}$~~~~~ \\
$6$~~~~ & $4.0007\times 10^{-3}$~~~~~~  & $4.001\times 10^{-3}$~~~~~ \\
\end{tabular}
\caption{Moments of the distribution $P(\widetilde{{\Omega}})$. Numerical
values are computed for the distribution in Fig.~1.}
\end{table}


\begin{references} 
\bibitem{mont} H. L. Montgomery, {\sl Proc. Symp. Pure Math.} {\bf 24}, 181
 (1973); D. Goldston and H. L. Montgomery, {\sl Proceeding Conf. at
 Oklahoma State Univ. 1984}, edited by A. C. Adolphson et al, p.183.
\bibitem{odlyzko} A. M. Odlyzko, {``The $10^{20}$-th zero of the Riemann zeta
 function and $175$ million of its neighbors''}, AT \& T Report, 1992
 (unpublished); see also http://www.research.att.com/\verb+~amo+/.
\bibitem{bgs} O. Bohigas, M.-J. Giannoni, in {\sl Mathematical and
 Computational Methods in Nuclear Physics} edited by J. S. Dehesa et al, 
 {\it Lectures Notes in Physics} {\bf 209}, (Springer Verlag, Berlin, 1984)
 p.1.
\bibitem{berry} M. V. Berry, in {\sl Quantum Chaos and Statistical Nuclear
  Physics} edited by T. H. Seligman and H. Nishioka, {\it Lectures Notes in
  Physics} {\bf 263}, (Springer Verlag, Berlin, 1986) p.1. 
\bibitem{sar} P. Sarnak, {\sl Curr. Dev. Math.}, edited by R. Bott et al.
 (1997) p.84.
\bibitem{bk} M. V. Berry and J. P. Keating, {\sl SIAM Review} {\bf 41}, 236 
 (1999).
\bibitem{bls1} the sum of complex zeros in small energy windows located at a
 height $E$ was considered in O. Bohigas, P. Leboeuf and M. J. S\'anchez,
 {\sl Physica D} {\bf 131}, 186 (1999).
\bibitem{tit} E. C. Titchmarsh, {\sl The Theory of the Riemann Zeta Function}
 (Oxford, Clarendon Press, 1986).
\bibitem{gutz} M. C. Gutzwiller, {\sl Chaos in Classical and Quantum 
 Mechanics} (Springer--Verlag, New York, 1990). 
\bibitem{selberg} A. Selberg, {\sl Collected Papers}, Vol. {\bf I}, 
 (Springer Verlag, Berlin, 1989) p.214.
\bibitem{lm} P. Leboeuf and A. Monastra, {\sl Phys. Rev. B} {\bf 62}, 12617
 (2000).
\bibitem{kzs} N. M. Katz and P. Sarnak, {\sl Bull. Amer. Math. Soc.} {\bf 36},
 1 (1999).
\bibitem{hkssz} F. Haake, M. Ku\'s, H.-J. Sommers, H. Schomerus and K.
 Zyczkowski, {\sl J. Phys. A} {\bf 29}, 3641 (1996).
\bibitem{ks} J. P. Keating and N. C. Snaith, {\sl Comm. Math. Phys.} {\bf 214},
 57 (2000).
\bibitem{tsang} K.-M. Tsang, {\sl Proc. London Math. Soc.} {\bf 65}, 65 (1992).
\bibitem{bdl} P. M. Bleher, F. J. Dyson, and J. L. Lebowitz, {\sl Phys. Rev.
 Lett.} {\bf 71}, 3047 (1993).
\end{references}
\end{document}